\documentclass[preprint,showpacs,showkeys,preprintnumbers,amsmath,amssymb,superscriptaddress]{revtex4}

\usepackage{graphicx}
\usepackage{dcolumn}
\usepackage{bm}
\usepackage[latin1]{inputenc}

\begin{document}

\author{J. G. Oliveira Jr.}
\affiliation{Centro de Forma\c{c}\~{a}o de Professores, Universidade
Federal do Rec\^{o}ncavo da Bahia, 45.300-000, Amargosa, BA, Brazil}
\affiliation{Departamento de F\'{\i}sica, Instituto de Ciências
Exatas, Universidade Federal de Minas Gerais, C.P. 702, 30161-970,
Belo Horizonte, MG, Brazil}
\author{R. Rossi Jr.}
\affiliation{Departamento de F\'{\i}sica, Instituto de Ciências
Exatas, Universidade Federal de Minas Gerais, C.P. 702, 30161-970,
Belo Horizonte, MG, Brazil}
\author{M. C. Nemes}
\affiliation{Departamento de F\'{\i}sica, Instituto de Ciências
Exatas, Universidade Federal de Minas Gerais, C.P. 702, 30161-970,
Belo Horizonte, MG, Brazil}

\title{Multipartite entanglement control via Quantum Zeno Effect}

\begin{abstract}
We develop a protocol based on $2M$ pairwise interacting qubits,
which through Quantum Zeno Effect controls the entanglement
distribution of the system. We also show that if the coupling
constants are different the QZE may be used to achieve perfect
entanglement swap.
\end{abstract}

\maketitle

\emph{Introduction}

Entangled states as Quantum Mechanics was constructed were in the
heart of feverous debates about apparent paradoxes such as EPR
(Einstein, Podolsky e Rosen )\cite{art1}. Nowadays the status of
entangled states changed radically from Gedanken experiments to
actual tools for various technological improvements, such as in the
area of quantum computing \cite{art2}. From the point of view of
realizing quantum computation one of the most difficult and
important tasks is the entanglement control of a large series of
systems.

Recently a system composed by four qubits interacting pairwise has
been exhaustively studied. Many important features of entanglement
dynamics were clarified by this analytic model, namely: sudden death
of entanglement \cite{art3,art4}, relation between energy and
entanglement \cite{art5}, pairwise concurrence dynamics \cite{art6},
entanglement invariant for this model \cite{art7} and entanglement
protection \cite{art8}.

In the present contribution we extend that study to a system
composed of $M$ pairs of qubits interacting pairwise, whose
entanglement distribution control is performed via the QZE (Quantum
Zeno Effect). We basically present two applications for the scheme:
a)We show that systems of four qubits whose autonomous dynamics have
different coupling constants may be slowed down by QZE in such a way
that the swap is completed as if the coupling constants were the
same. b) We show that generical entangled states in the $2M$ qubits
system can be transferred from any partition to any other directly,
independently of the distance between the partitions. This is
achieve, again through QZE.

\emph{The system}

Let us consider a multi-partite system composed by $2M$ qubits
interacting pairwise, with the hamiltonian given by

\begin{equation}
H=\sum_{k=1}^{M}H_{a_{k}A_{K}},\label{ham}
\end{equation}
where
\begin{equation}
H_{a_{k}A_{K}}=\hbar\omega_{a_{k}}\sigma_{z}^{a_{k}}
+\hbar\omega_{A_{k}}\sigma_{z}^{A_{k}}+\hbar
g_{k}\left(\sigma_{-}^{a_{k}}\sigma_{+}^{A_{k}}+\sigma_{+}^{a_{k}}\sigma_{-}^{A_{k}}\right),
\end{equation}
$g_{k}$ is the coupling constant of the $k$-th pair. To avoid
unnecessary complications let us consider
$\omega_{a_{k}}=\omega_{A_{k}}=\omega$. Notice that all the terms in
equation (\ref{ham}) commute with each other. Therefore, the global
system evolution can be separated in evolutions of each pair and
written as:

\begin{equation}
|\psi(T)\rangle=e^{-iHT/\hbar}|\psi(0)\rangle=\left(e^{-iH_{a_{1}A_{1}}T/\hbar}\otimes
e^{-iH_{a_{2}A_{2}}T/\hbar}\otimes \ldots\otimes
e^{-iH_{a_{M}A_{M}}T/\hbar}\right)|\psi(0)\rangle.
\end{equation}

If the initial state is an entangled state, the unitary evolution
will distribute the entanglement through the system. To give an
example of this entanglement dynamics suppose that
$|\psi(0)\rangle=\frac{1}{\sqrt{M}}\left(|1_{a_{1}}0_{a_{2}}\ldots
0_{a_{M}}\rangle+|0_{a_{1}}1_{a_{2}}\ldots
0_{a_{M}}\rangle+\ldots+|0_{a_{1}}0_{a_{2}}\ldots
1_{a_{M}}\rangle\right)|0_{A_{1}}0_{A_{2}}\ldots 0_{A_{M}}\rangle$
and $g_{j}=g$ (where $j=1,2\ldots M$). Initially there is a
maximally entangled state on the part that contains the qubits
$\{a_{k}\}$, and a factorized state on partition $\{A_{k}\}$. The
evolution of $|\psi(0)\rangle$ governed by the hamiltonian
(\ref{ham}) at time $T=\frac{\pi}{2g}$ results on an entanglement
swap between parts $a$ and $A$. Notice that the entanglement
distribution induced by the free evolution is dynamical and one has
not much control over it.

In this contribution, we present a protocol that allows for a more
incisive control of the entanglement distribution on the system. By
performing Zeno-like measurements on one component of any pair, it
is possible to inhibit the dynamics of this pair. Therefore, a
controlled evolution can be build up by $N$ steps of free
interactions followed by appropriate projections. For example, to
control the $j$-th pair of $|\psi(0)\rangle$ we must perform $N$
projective measurements on one of the qubits $a_{j}$ or $A_{j}$.
This controlled evolution can be written as:

\begin{equation}
\left(P_{j}e^{-iHT/N\hbar}\right)^{N}|\psi(0)\rangle=\left[e^{-iH_{a_{1}A_{1}}T/\hbar}
\otimes \ldots\otimes
\left(P_{j}e^{-iH_{a_{j}A_{j}}T/N\hbar}\right)^{N}\otimes
\ldots\otimes e^{-iH_{a_{M}A_{M}}T/\hbar}\right]|\psi(0)\rangle,
\end{equation}
where $P_{j}$ projects one of the qubits $a_{j}$ or $A_{j}$ in its
initial state, therefore by QZE the dynamics of the chosen pair is
inhibited when $N\rightarrow\infty$. Notice that $P_{j}$ acts only
on the subsystem of the j-th pair, so the evolution of all the other
pairs of qubits is free. This monitored evolution allow us to
control the entanglement distribution on the system, selecting the
pairs of qubits that will be free to evolve and the ones that will
have their evolution frozen.

\emph{Applications:}

\emph{a)Double Jaynes-Cummings Model}

In this section we present an explicit calculation for the time
evolution of two pairs of qubits when one of them is subjected to
$N$ projective measurements. The model for two qubits interacting
pairwise is refereed to as Double Jaynes-Cummings model and the
hamiltonian that governs the free evolution of this system is
(\ref{ham}) with $M=2$.

Let us consider the initial state

\begin{equation}
|\psi(0)\rangle=\left(\alpha_{0}|1_{a_{1}}0_{a_{2}}\rangle+\beta_{0}|0_{a_{1}}1_{a_{2}}\right)|0_{A_{1}},0_{A_{2}}\rangle,\label{est1}
\end{equation}

The free evolution of (\ref{est1}) and its entanglement dynamics
were studied in Ref.\cite{art5,art7}. An interesting aspect of this
free evolution is the probability of entanglement swap, i.e., when
$g_{1}=g_{2}$ the entangled state,initially prepared in one
partition is completely transferred to the other partition of the
system when the evolution time is $t=\frac{j\pi}{2g}$ ($j$ is an odd
number). The requirement for the coupling constants to be equal
($g_{1}=g_{2}$) may bring some difficulties for empirical
implementations of this entanglement swap. We show that if the
dynamics is controlled by QZE, the entanglement swap can be obtained
even with different coupling constants. To control the dynamics let
us introduce the projector $P_{2}=I_{a_{1}}\otimes I_{a_{2}}\otimes
I_{A_{1}}\otimes |0_{A_{2}}\rangle\langle 0_{A_{2}}|$, which acts on
the subsystem $A_{2}$ projecting in its initial state.

The vector state submitted to the controlled evolution (evolution
divided by $N$ projective measurements on $A_{2}$) is given by

\begin{equation}
|\psi(t)\rangle^{N}=\alpha_{0}e^{-iH_{a_{1}A_{1}}N\tau/\hbar}|1_{a_{1}}0_{A_{1}}\rangle\otimes|0_{a_{2}0_{A_{2}}}\rangle+
\beta_{0}|0_{a_{1}}0_{A_{1}}\rangle\otimes\left(P_{2}e^{-iH_{a_{2}A_{2}}\tau/\hbar}\right)^{N}|1_{a_{2}0_{A_{2}}}\rangle,
\end{equation}
where $N\tau=t$ and

\begin{eqnarray}
\left(P_{2}e^{-iH_{a_{2}A_{2}}\tau/\hbar}\right)^{N}&=&\left(\cos^{N}(g_{2}t)|1_{a_{2}}\rangle\langle
1_{a_{2}}|+|0_{a_{2}}\rangle\langle
0_{a_{2}}|\right)\otimes|0_{A_{2}}\rangle\langle
0_{A_{2}}|\\&&-i\tan(g_{2}t)\cos^{N}(g_{2}t)|1_{a_{2}}\rangle\langle
0_{a_{2}}|\otimes|0_{A_{2}}\rangle\langle 1_{A_{2}}|.\nonumber
\end{eqnarray}

The vector state after the controlled evolution can be written as:

\begin{equation}
|\psi(t)\rangle^{N}=\frac{1}{\sqrt{|\alpha_{0}|^{2}\left[1-\cos^{2N}(g_{2}\tau)\right]+\cos^{2N}(g_{2}\tau)}}\left(|\mu(t)\rangle
|0_{A_{1}},0_{A_{2}}\rangle-i|\nu(t)\rangle|1_{A_{1}},0_{A_{2}}\rangle\right),\label{vetorestado1}
\end{equation}
where

\begin{eqnarray}
|\mu(t)\rangle&=&\alpha_{0}\cos(g_{1}t)|1_{a_{1}},0_{a_{2}}\rangle+\beta_{0}\cos^{N}(g_{2}\tau)|0_{a_{1}},1_{a_{2}}\rangle, \\
|\nu(t)\rangle&=&\alpha_{0}\sin(g_{1}t)|0_{a_{1}},0_{a_{2}}\rangle.
\end{eqnarray}

Taking the limit $N\rightarrow\infty$ in eq. (\ref{vetorestado2}):

\begin{equation}
\lim_{N\rightarrow\infty}|\psi(t)\rangle^{N}=\left[\alpha_{0}\cos(g_{1}t)|1_{a_{1}},0_{a_{2}}\rangle+
\beta_{0}|0_{a_{1}},1_{a_{2}}\rangle\right]|0_{A_{1}},0_{A_{2}}\rangle-i\alpha_{0}\sin(g_{1}t)|0_{a_{1}}
,0_{a_{2}}\rangle|1_{A_{1}},0_{A_{2}}\rangle,\label{vetorestado2}
\end{equation}

To obtain the entanglement swap in this system, that has different
coupling constants for the pairs, we must inhibit, through QZE, the
excitation transfer of one pair for a certain period of time. If
$g_{1}<g_{2}$ ($g_{2}<g_{1}$) the excitation transfer in the pair
$a_{2}, A_{2}$ ($a_{1}, A_{1}$) is faster than the transfer in
$a_{1}, A_{1}$ ($a_{2}, A_{2}$). The transfer in the faster pair,
$a_{2}, A_{2}$ ($a_{1}, A_{1}$), must be inhibited for a period of
time given by
$T=\frac{\pi}{2}\left(\frac{1}{g_{1}}-\frac{1}{g_{2}}\right)$
$\left(T=\frac{\pi}{2}\left(\frac{1}{g_{2}}-\frac{1}{g_{1}}\right)\right)$.
Therefore, the evolution that allows for the entanglement swap is
composed by two parts. In the first part, which happens for the
period of time $T$, the evolution of one pair (the fastest pair) is
inhibited by QZE, while the other pair evolves freely. In the second
part of the evolution both pairs evolve freely. The total time of
evolution must correspond to a $\pi$ pulse for the slowest pair. The
fastest pair freezing in the first part of the total evolution
allows the complete excitation transfer in both pairs to take place
at the same exact time, this coincidence is an essential factor for
the entanglement swap.

Another interesting consequence of the partial control is that the
concurrence \cite{art9} for the qubits $a_{1}$ and $a_{2}$ after $N$
projective measurements (\ref{vetorestado1}), given by

\begin{equation}
C^{N}_{a_{1}a_{2}}(t)=\frac{2|\alpha_{0}\beta_{0}\cos(gt)\cos^{N}(g\tau)|}{|\alpha_{0}|^2+|\beta_{0}|^2\cos^{2N}(g\tau)},
\end{equation}
becomes, in the limit $N\rightarrow\infty$, identical to the
concurrence calculated in Ref.\cite{art10}, where the entanglement
dynamics between an isolated atom and a Jaynes-Cummings atom is
studied

\begin{equation}
\lim_{N\rightarrow\infty}C^{N}_{a_{1}a_{2}}(t)=2|\alpha_{0}\beta_{0}\cos(gt)|.
\end{equation}

Therefore using QZE, one can extract the entanglement dynamics of
the system studied in Ref.\cite{art10} from a double Jaynes-Cummings
system. We consider the initial state
$\left(\alpha_{0}|1_{a_{1}}0_{a_{2}}\rangle+\beta_{0}|0_{a_{1}}1_{a_{2}}\right)|0_{A_{1}},0_{A_{2}}\rangle$
in the calculation, but the same results can be shown for
$\left(\alpha_{0}|1_{a_{1}}1_{a_{2}}\rangle+\beta_{0}|0_{a_{1}}0_{a_{2}}\right)|0_{A_{1}},0_{A_{2}}\rangle$
as an initial state.

\emph{b) Transferring entangled states}

In this section we show how to transfer the entanglement from one
partition of the $2M$ qubits system interacting pairwise, to any
other partition of this system, using QZE and unitary evolution. For
simplicity let us consider in the calculation an eight qubits system
(the generalization for 2M qubits is straightforward). The eight
qubits are coupled pairwise and the time evolution is governed by
the Hamiltonian in equation (\ref{ham})(with $M=4$). The system is
prepared in the initial state

\begin{equation}
|\psi(0)\rangle=\left(|\phi^{+}\rangle+|\psi\rangle\right)\otimes|0_{A_{1}}0_{A_{2}}0_{A_{3}}0_{A_{4}}\rangle,
\end{equation}
where

\begin{eqnarray}
|\phi^{+}_{a_{1},a_{2},a_{3},a_{4}}\rangle&=&c_{1}|1_{a_{1}}1_{a_{2}}1_{a_{3}}1_{a_{4}}\rangle+c_{6}|0_{a_{1}}0_{a_{2}}0_{a_{3}}0_{a_{4}}\rangle,\\
|\psi_{a_{1},a_{2},a_{3},a_{4}}\rangle&=&c_{2}|1_{a_{1}}0_{a_{2}}0_{a_{3}}0_{a_{4}}\rangle+
c_{3}|0_{a_{1}}1_{a_{2}}0_{a_{3}}0_{a_{4}}\rangle+\\
&&c_{4}|0_{a_{1}}0_{a_{2}}1_{a_{3}}0_{a_{4}}\rangle+
c_{5}|0_{a_{1}}0_{a_{2}}0_{a_{3}}1_{a_{4}}\rangle,
\end{eqnarray}

Notice that a general entangled state is prepared in the partition
$\{a_{k}\}$. Now, suppose we want to transfer it to the partition
composed by the qubits $a_{3}-a_{4}-A_{1}-A_{2}$. The procedure is
simple, let the qubits $a_{1},a_{2},A_{1}$ and $A_{2}$ undergo a
$\pi$ pulse with time evolution governed by the Hamiltonian in
equation (\ref{ham}) (with $M=4$), inhibiting the evolution of
qubits $a_{3}-a_{4}-A_{3}-A_{4}$ by QZE. This dynamics gives us the
state

\begin{eqnarray*}
\left|\psi\left(\frac{\pi}{2g}\right)\right\rangle&=&-c_{1}|0_{a_{1}}0_{a_{2}}1_{a_{3}}1_{a_{4}}\rangle|1_{A_{1}}1_{A_{2}}0_{A_{3}}0_{A_{4}}\rangle+
c_{2}|0_{a_{1}}0_{a_{2}}0_{a_{3}}0_{a_{4}}\rangle|1_{A_{1}}0_{A_{2}}0_{A_{3}}0_{A_{4}}\rangle+\\
&&c_{3}|0_{a_{1}}0_{a_{2}}0_{a_{3}}0_{a_{4}}\rangle|0_{A_{1}}1_{A_{2}}0_{A_{3}}0_{A_{4}}\rangle+
c_{4}|0_{a_{1}}0_{a_{2}}1_{a_{3}}0_{a_{4}}\rangle|0_{A_{1}}0_{A_{2}}0_{A_{3}}0_{A_{4}}\rangle+\\
&&c_{5}|0_{a_{1}}0_{a_{2}}0_{a_{3}}1_{a_{4}}\rangle|0_{A_{1}}0_{A_{2}}0_{A_{3}}0_{A_{4}}\rangle
+c_{6}|0_{a_{1}}0_{a_{2}}0_{a_{3}}0_{a_{4}}\rangle|0_{A_{1}}0_{A_{2}}0_{A_{3}}0_{A_{4}}\rangle,
\end{eqnarray*}
which can be written as:

\begin{eqnarray*}
\left|\psi\left(\frac{\pi}{2g}\right)\right\rangle&=&|0_{a_{1}}0_{a_{2}}\rangle\left(
|\phi^{-}_{a_{3},a_{4},A_{1},A_{2}}\rangle+|\psi_{a_{3},a_{4},A_{1},A_{2}}\rangle\right)|0_{A_{3}}0_{A_{4}}\rangle,
\end{eqnarray*}
where

\begin{eqnarray}
|\phi^{-}_{a_{3},a_{4},A_{1},A_{2}}\rangle&=&-c_{1}||1_{a_{3}}1_{a_{4}}1_{A_{1}}1_{A_{2}}\rangle+c_{6}|0_{a_{3}}0_{a_{4}}0_{A_{1}}0_{A_{2}}\rangle,\\
|\psi_{a_{3},a_{4},A_{1},A_{2}}\rangle&=&c_{2}|0_{a_{3}}0_{a_{4}}1_{A_{1}}0_{A_{2}}\rangle+
c_{3}|0_{a_{3}}0_{a_{4}}0_{A_{1}}1_{A_{2}}\rangle+\\
&&c_{4}|1_{a_{3}}0_{a_{4}}0_{A_{1}}0_{A_{2}}\rangle+
c_{5}|0_{a_{3}}1_{a_{4}}0_{A_{1}}0_{A_{2}}\rangle,
\end{eqnarray}

It is clear that we transfer an entangled state from the subsystem
$a_{1}-a_{2}-a_{3}-a_{4}$ to the subsystem
$a_{3}-a_{4}-A_{1}-A_{2}$.

In summary we have constructed a protocol for an extended system and
shown how the use of QZE may help control the entanglement
distribution and entanglement swap. A further study in this regard
is actually finding a system where this protocol may be implemented
and to discuss the role of deleterious environment effects which
will affect the number of effective qubits.

The authors acknowledge financial support by CNPq.

\end{document}